\documentclass[twocolumn,prl,showpacs,floatfix,altaffilletter]{revtex4}
\usepackage{graphicx}
\usepackage{bm}
\usepackage{amsmath}
\usepackage{sansmath}
\usepackage{amsfonts}
\usepackage{amssymb}
\usepackage{xcolor}
\begin{document}
\title{Non-Coplanar and Counter-Rotating Incommensurate Magnetic Order \\
Stabilized by Kitaev Interactions in $\bm{\gamma}$-Li$_2$IrO$_3$}
\author{A. Biffin$^1$, R.\ D. Johnson$^1$, I. Kimchi$^2$, R. Morris$^1$, A. Bombardi$^3$,
J.\ G. Analytis$^{2,4}$, A. Vishwanath$^{2,4}$, and R. Coldea$^1$}
\affiliation{$^{1}$Clarendon Laboratory, University of Oxford,
Parks Road, Oxford OX1 3PU, U.K.} \affiliation{$^{2}$Department of
Physics, University of California, Berkeley, California 94720.
USA.} \affiliation{$^{3}$Diamond Light Source Ltd., Didcot OX11
0DE, U.K.} \affiliation{$^{4}$Materials Science Division, Lawrence
Berkeley National Laboratory, Berkeley, California 94720. USA.}

\begin{abstract}
Materials that realize Kitaev spin models with bond-dependent
anisotropic interactions have long been searched for, as the
resulting frustration effects are predicted to stabilize novel
forms of magnetic order or quantum spin liquids. Here we explore
the magnetism of $\gamma$-Li$_2$IrO$_3$, which has the topology of
a 3D Kitaev lattice of inter-connected Ir honeycombs. Using
resonant magnetic x-ray diffraction we find a complex, yet
highly-symmetric incommensurate magnetic structure with
non-coplanar and counter-rotating Ir moments. We propose a minimal
Kitaev-Heisenberg Hamiltonian that naturally accounts for all key
features of the observed magnetic structure. Our results provide
strong evidence that $\gamma$-Li$_2$IrO$_3$ realizes a spin
Hamiltonian with dominant Kitaev interactions.
\end{abstract} \maketitle

Magnetic materials with bond-dependent anisotropic interactions
are candidates to display novel forms of magnetic order or quantum
spin liquid states, as exemplified by the Kitaev model on the
honeycomb lattice \cite{kitaev}. Here all spins interact via
nearest-neighbor Ising exchanges, but a different Ising axis
($\mathsf{x},\mathsf{y},\mathsf{z}$) applies for the three
different bonds emerging out of each lattice site. This leads to
strong frustration effects that stabilize a novel gapless quantum
spin liquid state with exotic excitations (Majorana fermions),
which is exactly solvable in two dimensions. It was theoretically
proposed \cite{chaloupka} that such exotic Hamiltonians might be
realized in magnetic materials containing edge-sharing cubic
IrO$_6$ octahedra. The magnetic ground state of Ir$^{4+}$
including the cubic crystal field and spin-orbit coupling is a
complex spin-orbital doublet with $J_{\rm eff}=1/2$ \cite{kim},
and super-exchange through the two 90$^{\circ}$ Ir-O-Ir paths is
expected to lead to a dominant Ising interaction for the moment
components normal to the Ir-O$_2$-Ir plane \cite{chaloupka}. For a
three-fold coordinated IrO$_6$ octahedron this leads to
perpendicular Ising axes for the three nearest-neighbor bonds, as
required for a Kitaev model. The 2D honeycomb-lattice
$\alpha$-Na$_2$IrO$_3$ \cite{singh,liu,choi,ye,gretarsson} and
$\alpha$-Li$_2$IrO$_3$ \cite{omalley,singh-manni} are being
intensively explored as candidate Kitaev materials, but as yet no
clear evidence for novel Kitaev physics has been observed.

Generalizations of the Kitaev model to 3D lattices are also
expected to have quantum spin liquid states \cite{mandal,
lee,kimchi}. The recently-synthesized structural polytypes
``hyper-honeycomb" $\beta-$Li$_2$IrO$_3$ \cite{beta} and
``harmonic" honeycomb $\gamma-$Li$_2$IrO$_3$ \cite{modic}, which
maintain the local three-fold coordination of edge-sharing IrO$_6$
octahedra, are prime candidates to display 3D Kitaev physics. To
test for signatures of such physics we have performed resonant
magnetic x-ray diffraction (RMXD) measurements \cite{hill} on
single crystals of $\gamma$-Li$_2$IrO$_3$, scattering at the
strong Ir L$_3$ resonance \cite{liu}. We have determined the
complete magnetic structure for all 16 iridium sites in the unit
cell, and found an unexpectedly complex, yet highly symmetric
magnetic structure comprised of non-coplanar, counter-rotating
iridium magnetic moments located in zig-zag chains. Remarkably,
the magnetic structure exhibits no net ferromagnetic or
antiferromagnetic spin correlations, and as such one can rule out
a model Hamiltonian whose primary ingredient is the
nearest-neighbor Heisenberg interaction. Instead, motivated by the
work of Jackeli and Khaliullin \cite{jackeli}, and by arguments
based on susceptibility anisotropy \cite{modic, kimchi}, we
present a minimal spin Hamiltonian with dominant Kitaev
interactions that naturally reproduces all key features of the
observed magnetic order, in particular, we point out that counter
rotating spirals on the zig-zag chains are naturally generated by
Kitaev interactions. Our results therefore provide strong evidence
that dominant Kitaev couplings govern the magnetic interactions in
$\gamma$-Li$_2$IrO$_3$.

\begin{figure}[htbp]
\includegraphics[width=0.46\textwidth]{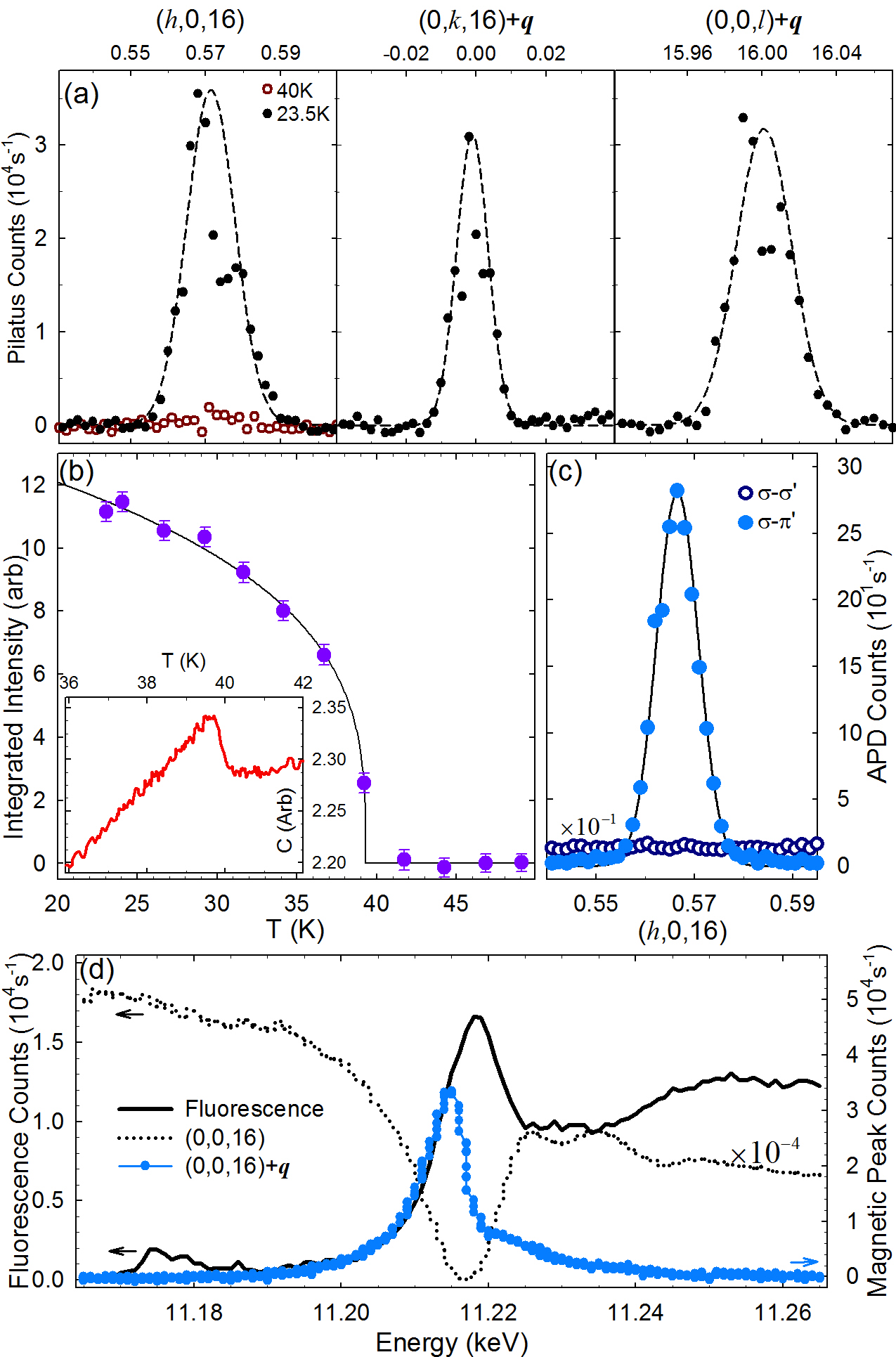}
\caption{\label{fig:braggpeak} (color online) Magnetic Bragg peak
at $(0,0,16)$$+$${\bm q}$. (a) Scans along orthogonal directions
in reciprocal space (filled/open symbols at base temperature/above
$T_{\rm N}$). Dashed lines are fits to a Gaussian shape. (b)
Temperature-dependence of the integrated magnetic peak intensity
(solid line is guide to the eye, temperature values are corrected
for beam heating effects, see text). Inset: specific heat data
showing an anomaly at the onset of magnetic order. (c) Scans with
a polarizer in the scattered beam: the magnetic signal is present
only in the $\sigma$-$\pi'$ channel (filled circles) and
disappears in the $\sigma$-$\sigma'$ (open circles) dominated by
charge scattering (intensity scaled by $1/10$). (d) Energy scan
through the magnetic peak (blue squares) and a structural Bragg
peak ($0,0,16$) (dotted line, scaled by $1/10^4$), as well as the
fluorescence scan (solid line).}
\end{figure}
\begin{figure}[htbp]
\includegraphics[width=0.46\textwidth]{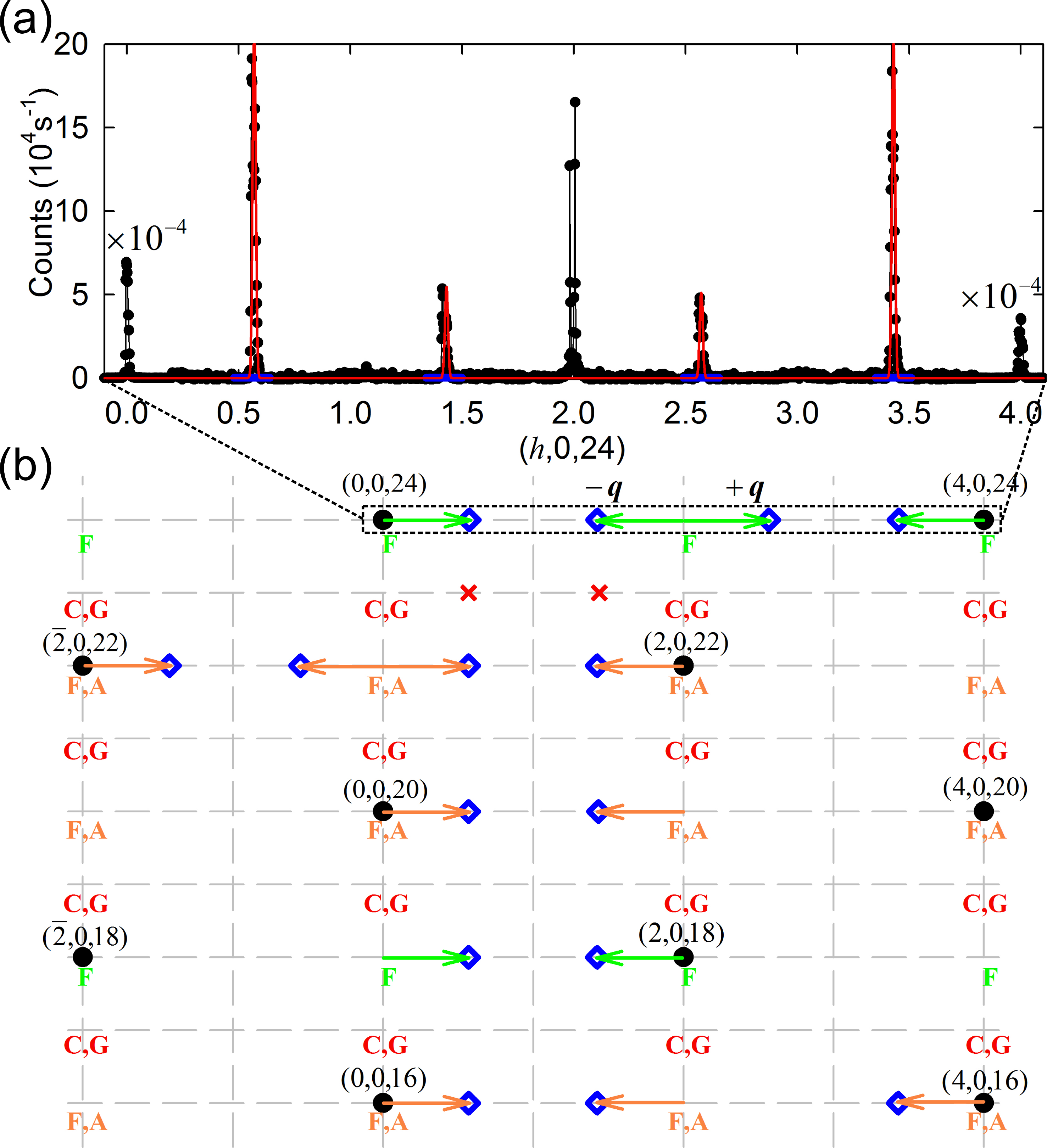}
\caption{\label{fig:hscan} (color online) (a) Scan along the
$(h,0,24)$ direction observing structural peaks at $h=0,4$
(intensity scaled by $1/10^4$ for clarity), a multiple-scattering
signal centered at $h=2$, and magnetic peaks at $h = 0+q,2\pm q,
4-q$. Solid red line is the calculated magnetic scattering
intensity \cite{magnetix} for the magnetic structure model
depicted in Fig.~\ref{fig:magstruct}. (b) $(h0l)$ reciprocal plane
with filled circles, diamonds and red crosses indicating positions
of structural peaks, measured magnetic peaks and the absence of
peaks, respectively. Lattice points are also labelled by the
magnetic basis vectors that have finite structure factor for
magnetic peaks at satellite $\pm{\bm q}$ positions.}
\end{figure}

The RMXD experiments were performed using the I16 beamline at
Diamond (see \cite{supplemental} for details). Systematic searches
along high-symmetry directions in reciprocal space revealed that
at low temperatures new magnetic Bragg peaks appeared at satellite
positions of reciprocal lattice points with an incommensurate
propagation vector $\bm{q}=(0.57(1),0,0)$ \cite{error}. The
satellite peaks were found to be as sharp as structural peaks in
all three reciprocal space directions, as illustrated for the
$(0,0,16)$$+$$\bm{q}$ reflection in Fig.~\ref{fig:braggpeak}a);
indicating coherent, 3D magnetic ordering. The peaks disappeared
upon heating [Fig.~\ref{fig:braggpeak}(a), open circles] and the
temperature-dependence of the intensity had a typical order
parameter behavior [see Fig.~\ref{fig:braggpeak}(b)]. The absolute
temperature values have been corrected for beam-heating effects
through a calibration against specific heat measurements on the
{\em same} sample, shown in Fig. \ref{fig:braggpeak}(b) inset,
which give $T_{\rm N}=39.5$\ K.

The magnetic origin of the satellite reflections was further
confirmed by analyzing the polarization of the scattered beam.
Fig.~\ref{fig:braggpeak}(c) shows that the peak at $(0,0,16)$$+$$
\bm{q}$ appeared only in the $\sigma$-$\pi'$ channel (filled
circles), and is absent in the $\sigma$-$\sigma'$ channel (open
circles), as expected for resonant diffraction that is of pure
magnetic origin \cite{hill}. An energy scan performed whilst
centered on the magnetic peak [Fig.~\ref{fig:braggpeak}(d)] showed
a large resonant enhancement of the scattered intensity, again as
expected for RMXD. The energy dependence is in stark contrast to
that characteristic of a nearby structural peak (dotted line).
Furthermore, the obtained resonance energy is similar to values
found in other iridates \cite{liu, boss} and agrees well with the
edge of the measured fluorescence signal from the sample (solid
line in Fig.~\ref{fig:braggpeak}(d)).

In total over 30 magnetic Bragg peaks were observed, and those
measured in the $(h0l)$-plane are labelled in
Fig.~\ref{fig:hscan}b). A representative scan along the ($h,0,24$)
direction is plotted in Fig.~\ref{fig:hscan}a), which shows strong
structural Bragg peaks centered at $h=0,4$, a multiple scattering
signal centered at $h=2$, and four magnetic Bragg peaks
symmetrically displaced away from the above reflections. The scan
illustrates the highly symmetric nature of the magnetic peak
intensities and that $\bm{q}$ is distinctly different from the
commensurate wavevector ($\frac{1}{2}00$).

\begin{figure}[t]
\includegraphics[width=0.46\textwidth]{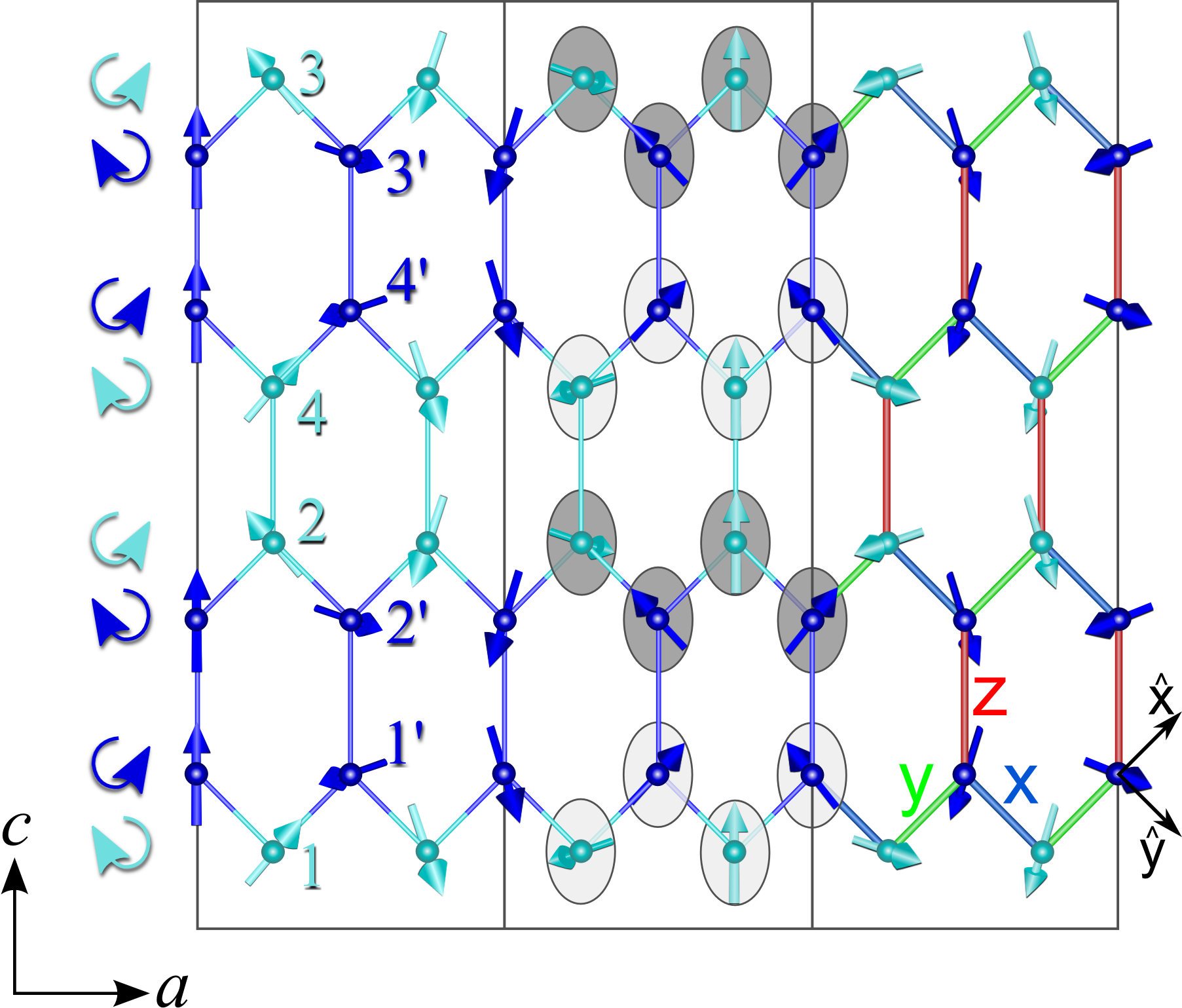}
\caption{\label{fig:magstruct} (color online) Projection of the
magnetic structure on the $(\bm{a},\bm{c})$ plane showing 3 unit
cells along the horizontal propagation direction $a$. Light and
dark blue arrows show the moments on the Ir and Ir$'$ sublattices,
with sites $1-4$ and $1'-4'$, respectively. Curly arrows on the
left side illustrate counter-rotating magnetic order between
consecutive sites along $c$. In unit cell 2 light ($-\phi$) and
dark ($+\phi$) shaded elliptical envelopes emphasize the
confinement of the moments to alternate planes obtained from the
$(ac)$ plane by a rotation by $\mp \phi$ around $c$. In unit cell
3 color of bonds indicates the anisotropy axis of the Kitaev
exchanges in (\ref{eq:ham}), with
$\eta=\mathsf{x},\mathsf{y},\mathsf{z}$ for blue/green/red bonds,
where $\mathsfbf{\hat{x}} =(\bm{\hat{a}}+\bm{\hat{c}})/\sqrt{2}$,
$\mathsfbf{\hat{y}} =(\bm{\hat{a}}-\bm{\hat{c}})/\sqrt{2}$ and
$\mathsfbf{\hat{z}} =\bm{\hat{b}}$).}
\end{figure}

The magnetic iridium ions are located on two inequivalent
sublattices in the orthorhombic unit cell, referred to as Ir and
Ir$'$, respectively (light and dark balls in
Fig.~\ref{fig:magstruct}). Each sublattice contains four sites in
the primitive cell labelled 1 to 4 and 1$'$ to 4$'$, respectively.
For a propagation vector ${\bm q}=(q,0,0)$ symmetry analysis
\cite{basireps} gives four types of magnetic basis vectors for
{\em each} of the two sublattices: $++++$($F$), $++--$($C$),
$+--+$($A$) and $+-+-$($G$) where the $\pm$ signs denote a
symmetry-imposed relation between the magnetic Fourier components
at the sites 1-4 and 1$'$-4$'$. There are no symmetry constraints
between the basis vectors on the two sublattices.

\begin{figure}[t]
\includegraphics[width=0.46\textwidth]{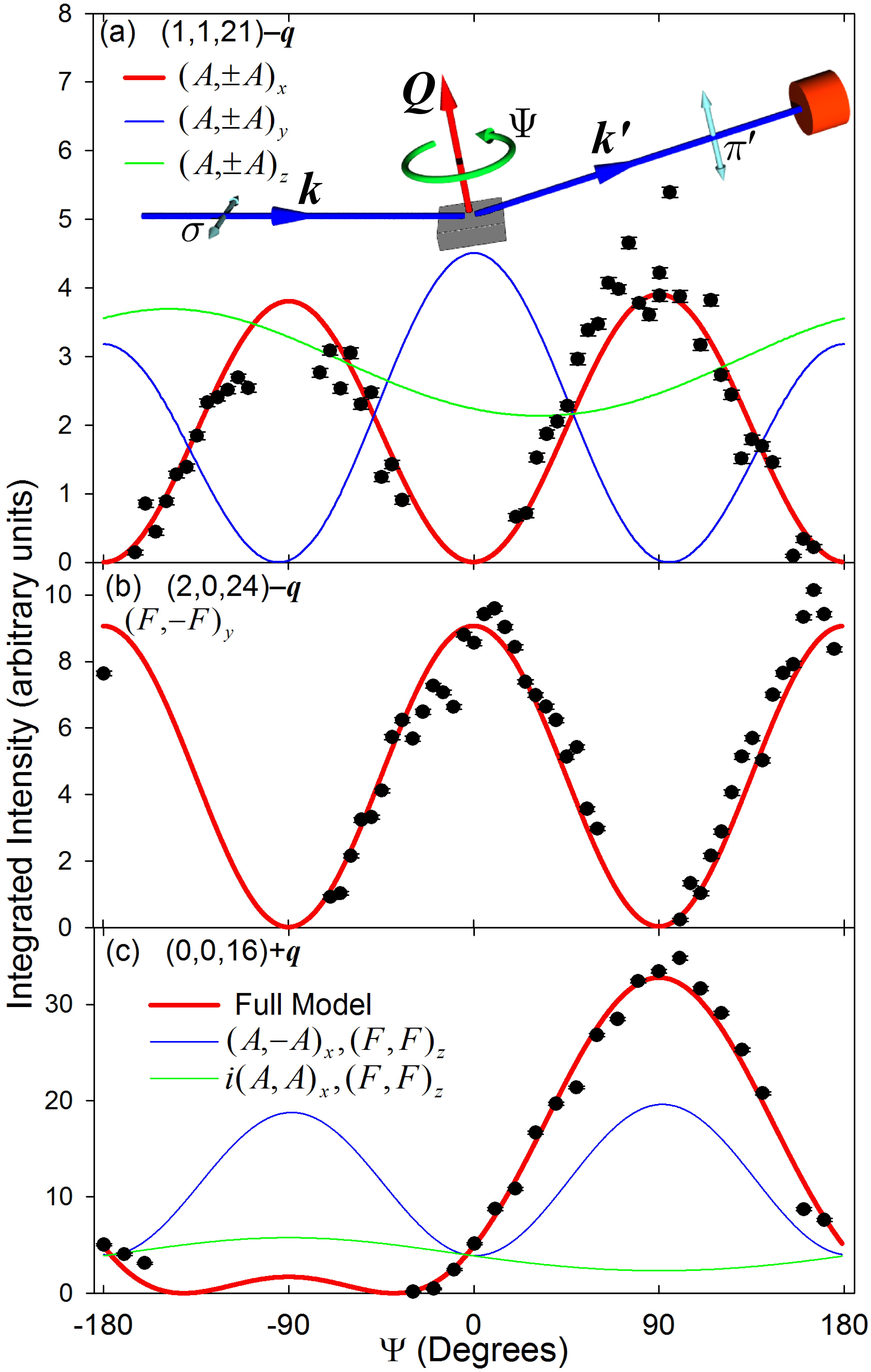}
\caption{\label{fig:azimuths} (color online) Integrated intensity
as a function of azimuth for three magnetic Bragg peaks, a)
pure-$A_x$, b) pure-$F_y$ and c) mixed-$F_zA_x$. Top diagram
illustrates the scattering geometry. Data points (filled circles)
are integrated peak intensities from rocking curve scans corrected
for absorption and Lorentz factor. Thick (red) lines show fits
that include all contributions to the RMXD structure factor
\cite{hill,magnetix} from the magnetic structure model
$i(A,-A)_x,-i(F,-F)_y,(F,F)_z$, depicted in
Fig.~\ref{fig:magstruct}. Blue/green curves in a,c) illustrate
that other phase combinations of basis vectors are ruled out.}
\end{figure}

Each of the four types of basis vectors has its own selection
rules for a non-zero structure factor, so their presence can be
directly confirmed from the observation of magnetic reflections at
certain positions, and in some cases one can also identify the
phase relation between the two sublattices. For example, all
magnetic peaks along the $(h,0,24)$ line in Fig.~\ref{fig:hscan}a)
can be uniquely assigned to scattering from $F$-type basis
vectors. Satellites at $h=0+q$ and $4-q$ arise from components
that are equal in magnitude and in phase on the two sublattices,
$(F,F)$ in short-hand notation, whereas the satellites at $h=2 \pm
q$ originate from scattering by components equal in magnitude, but
with opposite sign on the two sublattices, i.e. $(F,-F)$ (see
\cite{supplemental} for details). The overall selection rules for
magnetic scattering are illustrated in Fig.~\ref{fig:hscan}b). We
have ruled out the presence of both $C$ and $G$ basis vectors as
systematic searches (at 4 different azimuth angles) at the
satellite positions ($0,0,23)$$+$$\bm{q}$ and
($2,0,23)$$-$$\bm{q}$ (red crosses) gave no sign of a magnetic
signal. Furthermore, the observation of an $AG$ magnetic peak at
$(1,1,21)$$-$$\bm{q}$, $G$ being ruled out, confirms the presence
of an $A$ basis vector (azimuth scan in
Fig.~\ref{fig:azimuths}(a)).

The polarization dependence of the RMXD intensity allows a direct
identification of the orientation of the magnetic moments. For a
$\sigma$-polarized incident beam only the projection of the
magnetic moments along the scattered beam direction,
$\bm{\hat{k'}}$, contribute to the diffraction
intensity.\cite{hill} By rotating the sample around the scattering
vector ${\bm Q}=\bm{k'}-\bm{k}$ by the azimuth angle, $\Psi$, [see
diagram in Fig.\ref{fig:azimuths}a) inset] the projection of the
magnetic moments onto $\bm{\hat{k'}}$ changes, giving a clear
signature of the moment direction. We have measured the azimuth
dependence for three magnetic peaks close to the sample surface
normal, such that the $\Psi$ rotation is almost around (001). The
origin, $\Psi=0$, is defined as when the (010) direction is in the
scattering plane and pointing away from the source.
Fig.~\ref{fig:azimuths}a) shows the azimuth scan for a pure-$A$
magnetic Bragg peak. The intensity drops to zero at $\Psi=0$ and
$180^{\circ}$ and has maxima at $\pm90^{\circ}$, uniquely
identifying that scattering comes from magnetic moment components
along $x$ (here $x$, $y$, $z$ are along the orthorhombic $a$, $b$,
$c$ axes and scattering from $y$- and $z$-moment components, blue
and green lines, respectively, have been calculated for
comparison); hence identifying basis vector components in the
combination $(A,\pm A)_x$, where the two sublattices are assumed
to have equal magnitude moments. Similarly, the azimuth of the
pure-$F$ peak in Fig.~\ref{fig:azimuths}b) originates from
$y$-components antiparallel on the two sublattices, identifying
the basis vector $(F,-F)_y$. Fig.~\ref{fig:azimuths}c) shows the
azimuthal dependence for a mixed $FA$ peak, which uniquely
identifies it as coming from basis vector components $\pi/2$
out-of-phase in the combination $i(A,-A)_x, (F,F)_z$. We note that
this combination of relative phases between the $x$ and $z$
components on all the iridium sites is unique, where other
combinations can be qualitatively ruled out (see blue/green curves
in the same figure). The observed phase combination describes
counter-rotating moments between consecutive sites along $c$
(curly arrows in Fig.~\ref{fig:magstruct}), which form
counter-rotating zig-zag chains along $a$.

To determine the relative magnitudes of the magnetic moment
components we performed a simultaneous fit to the magnetic
scattering intensities in the three azimuth scans in
Fig.~\ref{fig:azimuths} with four free parameters: the magnitudes
of the moment amplitudes $M_x$ and $M_y$ relative to $M_z$, an
overall intensity scale factor for the $(1,1,21)$$-$$\bm{q}$ and
$(2,0,24)$$-$$\bm{q}$ peaks and a separate intensity scale factor
for the $(0,0,16)$$+$$\bm{q}$ peak (which was measured on the same
sample, but in a different experiment). The fit is shown by red
solid lines in Fig.~\ref{fig:azimuths}a-c), and gave values for
the moment magnitude ratios $M_x:M_y:M_z=0.65(4):0.58(1):1$. We
note that this also quantitatively reproduces the observed ratio
of the magnetic peak intensities in Fig.~\ref{fig:hscan}a) (red
line).

Imposing the constraint of near-constant magnitude moment at every
site requires the phase offset between the $x$ and $y$ components
to be $\pi$ or $0$, giving the basis vector combination
$i(A,-A)_x, i(-1)^m(F,-F)_y,(F,F)_z$, with $m=1$ or 2. Both give
similar structures and we plot in Fig.~\ref{fig:magstruct} the
case $m=1$. The moments are confined to rotate in one of two
planes, obtained from the ($ac$) plane by rotation around the
$c$-axis by an angle $\pm\phi$, with $\phi={\rm
tan}^{-1}\frac{M_y}{M_x}=42(2)^{\circ}$. The pattern is such that
neighboring iridium zig-zag chains have alternate orientations of
the spin rotation plane as indicated by the light and dark shaded
envelopes in Fig.~\ref{fig:magstruct}. The $m=2$ case simply gives
the opposite alternation of the rotation planes.

A key feature of the magnetic structure is the counter-rotation of
neighboring moments. On two such sites, say $1$ and $1'$, the
spins projected to the $ac$-plane are  $ \bm{S}_{1,1'}(\bm{r}) =
\bm{\hat{c}}\langle S^c \rangle \cos \bm{q}\cdot\bm{r} \pm
\bm{\hat{a}} \langle S^a \rangle \sin \bm{q}\cdot\bm{r} $. We now
rotate from the crystallographic $a,b,c$-axes to the Kitaev
$\mathsf{x},\mathsf{y},\mathsf{z}$-axes (see Fig.\
\ref{fig:magstruct} caption) and consider the correlation between
the $S^{\mathsf{x}}$ spin components $S_1^\mathsf{x}
S_{1'}^\mathsf{x}$  across an $\mathsf{x}$-type bond, or
$S_1^\mathsf{y} S_{1'}^\mathsf{y}$ across a $\mathsf{y}$-type
bond. The net averaged correlation is finite, $\left\langle
S_1^\mathsf{x} S_{1'}^\mathsf{x} \right\rangle_{\mathsf{x}}
=\left\langle S_1^\mathsf{y} S_{1'}^\mathsf{y}
\right\rangle_{\mathsf{y}} = \langle S^a \rangle \langle S^c
\rangle \tfrac{1}{2} \sin \tfrac{\pi q}{2}$. We see that along
each $\mathsf{x}$-type bond the spins are aligned when they point
along $\mathsf{x}$, and anti-aligned when they point along
$\mathsf{y}$, and similarly for $\mathsf{y}$-type bonds. Thus
Kitaev interactions can stabilize the counter-rotating moments
with a propagation vector $\bm{q}$ along $\bm{a}$. We therefore
construct the following Kitaev-Heisenberg Hamiltonian as a minimal
model
\begin{align}
&{\mathcal{H}} = \sum_{c-{\rm bonds}} \left[K_c S_i^{\eta_{ij}}
S_j^{\eta_{ij}} +
J_c \bm{S}_i\cdot \bm{S}_j + I_c^c S_i^c S_j^c \right] + \nonumber \\
&\sum_{d-{\rm bonds}}\!\!\left[K_d S_i^{\eta_{ij}} S_j^{\eta_{ij}}
+ J_d \bm{S}_i\cdot \bm{S}_j\right]  + \!\!\sum_{2^{\rm nd} \
\langle\langle i j \rangle\rangle} \!\! J_2 \bm{S}_i\cdot \bm{S}_j
\label{eq:ham}
\end{align}
where interactions along the vertical (along $c$) bonds are
denoted by the subscript $c$ and interactions along the zig-zag
(diagonal) bonds are denoted by the subscript $d$. $K_c$ and $K_d$
are the Kitaev interactions along $c$-bonds (of type
$\eta_{ij}=\mathsf{z}$) and $d$-bonds (of type
$\eta_{ij}=\mathsf{x}$ or $\mathsf{y}$), respectively. To prevent
$(0,0,q_c)$ instabilities we have introduced an Ising coupling
$I_c^c$ of the $S^c$ spin components, and finally a Heisenberg
coupling $J_2$ between second nearest neighbors. We take the
following values for the parameters (in units of meV): $K_c = -15,
K_d = -12, J_c = 5, J_d = 2.5,  I_c^c = -4.5, J_2 = -0.9 $
\cite{supplemental}, where the overall scale was set such as to
have the calculated ordering transition temperature agree with the
experimental value.

The Hamiltonian was analyzed in Fourier space using the
Luttinger-Tisza approximation \cite{supplemental}. This gave the
lowest-energy mode identical to the $(S^a,S^c)$ coplanar
projection of the magnetic structure in Fig.\ \ref{fig:magstruct}
with $\langle S^c \rangle > \langle S^a \rangle$. To obtain
fixed-length spins requires mixing with another mode, and the
lowest energy mode available at the same wavevector has collinear
order of the $S^b$ components with a pattern such that the mixed
mode exactly reproduce the observed non-coplanar structure.
Furthermore, the $S^b$ components are co-aligned along all the
$c$-axis bonds, and hence stabilized by the large FM $K_c$ Kitaev
exchange. The mixing amplitude, related to the tilt angle $\phi$,
is fixed for unit length spins, but changes continuously with the
Hamiltonian parameters. Decreasing the strength of the Kitaev
interactions prevents the ground state from producing unit-length
spins through this mixing mechanism, and importantly, we find that
the non-coplanar tilt angle observed in $\gamma$-Li$_2$IrO$_3$
requires relatively large Kitaev exchanges within the minimal
model.


To summarize, through RMXD measurements on $\gamma$-Li$_2$IrO$_3$
single crystals we have observed an incommensurate, non-coplanar
magnetic structure with counter-rotating moments. A
Kitaev-Heisenberg Hamiltonian can fully explain the observed
complex magnetic structure, providing strong evidence that
$\gamma$-Li$_2$IrO$_3$ is an experimental realization of 3D Kitaev
physics in the solid state.

This work was supported by EPSRC (UK) and by the U.S. Department
of Energy, Office of Basic Energy Sciences, Materials Sciences and
Engineering Division, under Contract No. DE-AC02- 05CH11231.

\vspace{2cm}

\makeatletter
\renewcommand{\thefigure}{S\@arabic\c@figure}
\renewcommand{\thetable}{S\@arabic\c@table}
\makeatother
\setcounter{figure}{0}

\section{Supplemental Material}

Here we provide additional information on 1) the crystal
structure, 2) the magnetic structure, 3) magnetic structure factor
calculations, 4) the sample and experimental setup used in the
resonant magnetic x-ray diffraction (RMXD) experiments, 5) the
RMXD intensity from moment-rotating structures, 6) a description
of how Kitaev interactions stabilize counter-rotating moments, and
7) details of the Luttinger-Tisza
analysis of the minimal model Hamiltonian.\\

\section{S1. Crystal Structure of $\bm{\gamma}$-L\MakeLowercase{i}$_2$I\MakeLowercase{r}O$_3$}

\begin{figure}[htbp]
\includegraphics[width=0.48\textwidth]{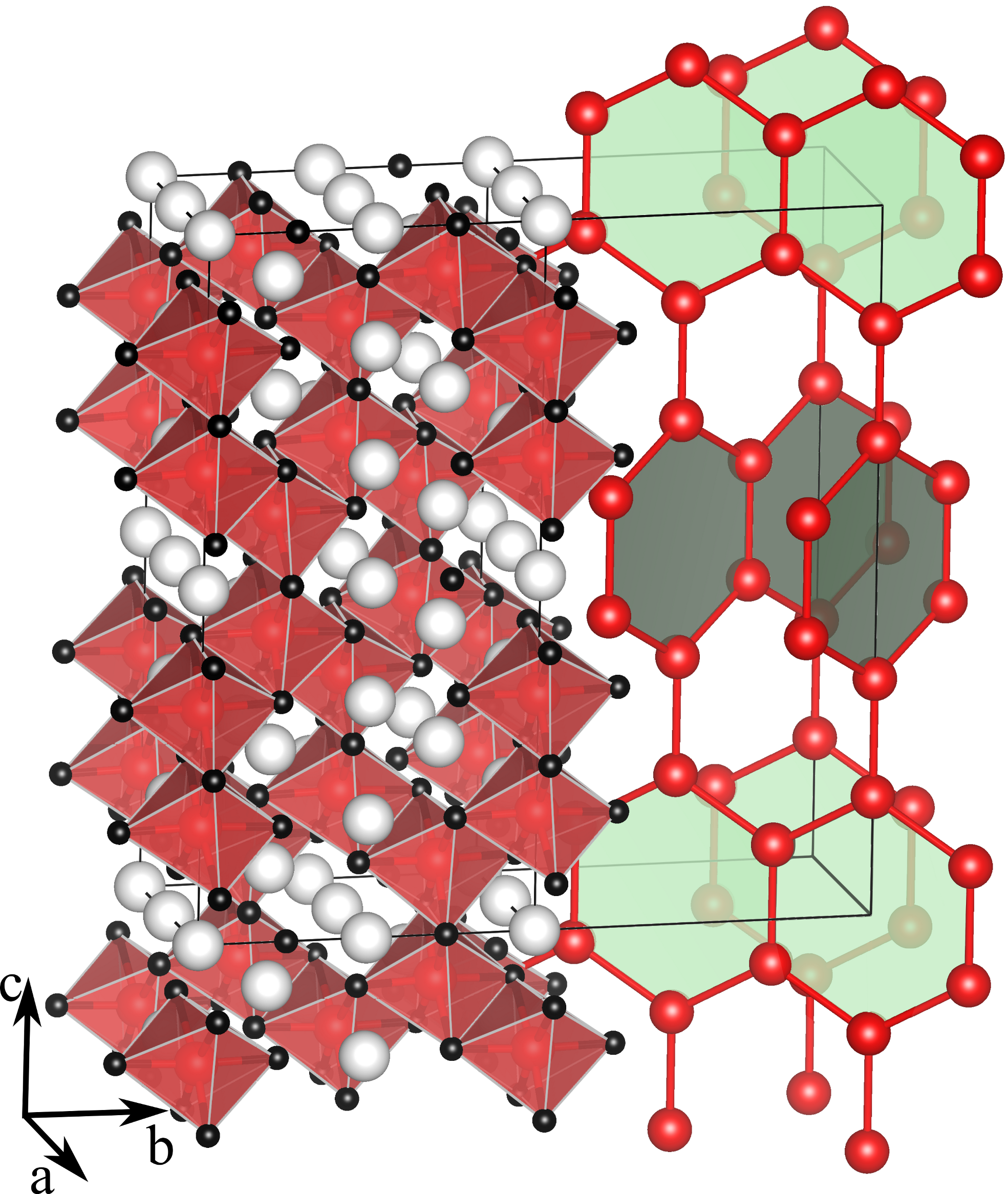}
\caption{(color online) Crystal structure of
$\gamma$-Li$_2$IrO$_3$. Two neighbouring unit cells are shown:
(left) full structure with Li (white balls), O (black) and  Ir
(red) located inside IrO$_6$ octahedra (shaded polyhedra), (right)
3D iridium lattice connectivity: honeycomb rows alternating in
orientation (light and dark shading) are interconnected along
$c$.} \label{fig:cccmstruct}
\end{figure}

$\gamma$-Li$_2$IrO$_3$ has an orthorhombic crystal structure
depicted in Fig.\ \ref{fig:cccmstruct} with edge-sharing IrO$_6$
octahedra arranged in a three-dimensional network with a
three-fold local coordination. The iridium atoms (red balls) form
vertically-linked honeycomb rows (light and dark shading) that run
alternatingly along the $\bm{a}\pm\bm{b}$ diagonals upon moving
along the $c$-axis. For reference the full structural parameters
from \cite{modic} are listed in Table\ \ref{tab:struc} (A.D.P. are
atomic displacement parameters). To simplify the notation for the
discussion of the magnetic structure we have labelled the two
iridium sublattices as Ir and Ir$'$.\\

\begin{table}[h!]
\caption{\label{tab:struc} Structural parameters of
$\gamma$-Li$_2$IrO$_3$ at 300\ K \cite{modic}.}
\begin{ruledtabular}
\begin{tabular}{c c c c c c}
\multicolumn{6}{l}{\textbf{Cell parameters}} \\
\multicolumn{6}{l}{Space group: $Cccm$ ($\#$66, origin choice 1)} \\
\multicolumn{6}{l}{Z = 16} \\
$a,b,c$(\AA): & 5.9119(3) & 8.4461(5) & 17.8363(10) \\
$\alpha,\beta,\gamma (^\circ)$: & 90 & 90 & 90 \\
\multicolumn{6}{l}{Volume (\AA$^3$): 890.61(9)}\\
\\
\multicolumn{6}{l}
{\textbf{Fractional atomic coordinates and isotropic A.D.P.'s}} \\
Atom & Site & $x$ & $y$ & $z$ & $U_\mathrm{iso}$(\AA$^2$)\\
\hline
 Ir & 8$k$ & 0.25 &  0.25 &  0.0836(2) &  0.0124(4) \\
 Ir$'$ & 8$i$ & 0.5 &  0.5 &  0.1670(3) &  0.0206(6) \\
 Li1 & 8$j$ & 0 &  0.5 &  0.3333 &  0.01 \\
 Li2 & 8$k$ & 0.75 &  0.25 &  0.25 &  0.01 \\
 Li3 & 8$k$ & 0.75 &  0.25 &  0.91667 &  0.01 \\
 Li4 & 4$c$ & 0.5 &  0.5 &  0.5 &  0.01 \\
 Li5 & 4$d$ & 0.5 &  0 &  0 &  0.01 \\
 O1 & 16$m$ & 0.77(1) &  0.515(3) &  0.087(4) &  0.02(1) \\
 O2 & 8$g$ &  0.72(2) &  0.5 &  0.25 &  0.04(1) \\
 O3 & 8$l$ & 0.00(1) &  0.262(8) & 0 &  0.006(9) \\
 O4 & 16$m$ &  0.49(1) &  0.262(6) &  0.163(3) &  0.006(9) \\
\end{tabular}
\end{ruledtabular}
\end{table}

\section{S2. Magnetic structure described in terms of basis vectors}

\begin{table}[tbh]
\caption{Irreducible representations and basis vectors for a
magnetic structure with propagation vector $\bm{q}=(q,0,0)$.}
\label{tab:irreps}
\par
\begin{center}
\begin{tabular}
[c]{l|c}\hline Irreducible & Basis vectors\\
representation & \\\hline
$\Gamma_1$ & $C_x,A_y,G_z$\\
$\Gamma_2$ & $F_x,G_y,A_z$\\
$\Gamma_3$ & $A_x,C_y,F_z$\\
$\Gamma_4$ & $G_x,F_y,C_z$\\
\end{tabular}
\end{center}
\end{table}

\begin{figure}[htbp]
\includegraphics[width=0.43\textwidth]{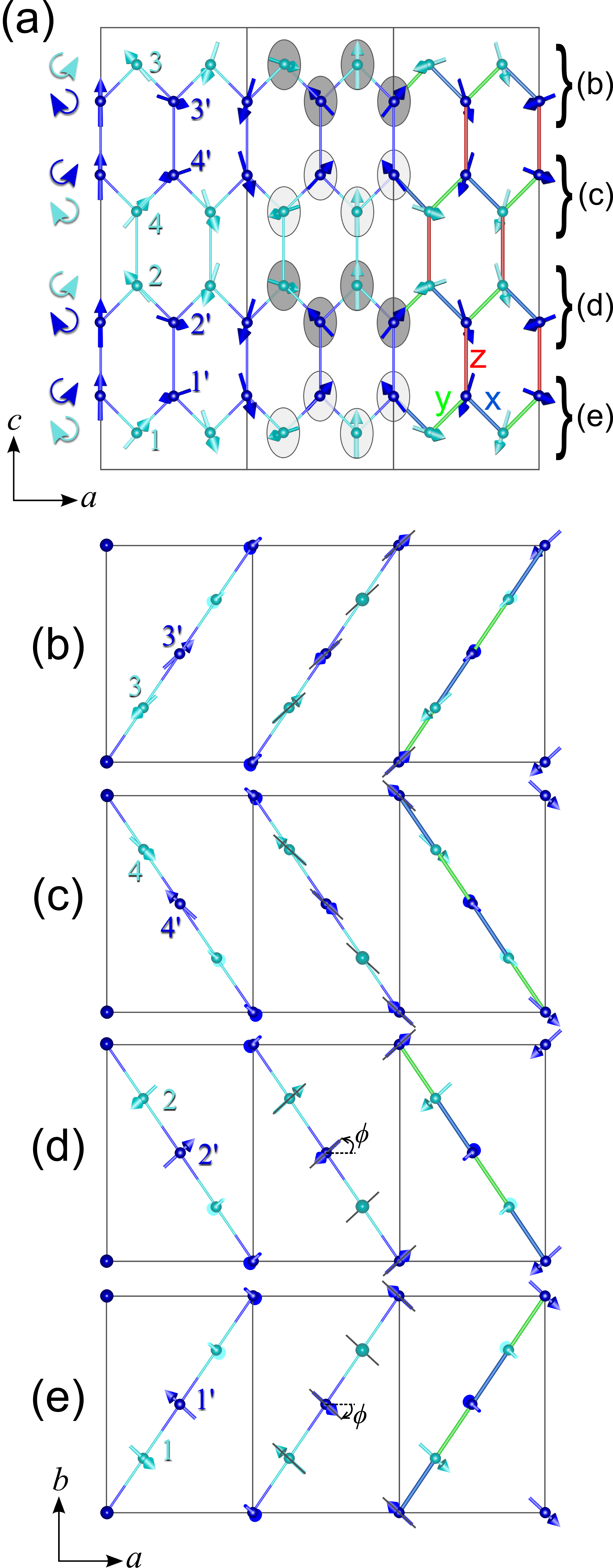}
\caption{(color online) (a) Projection of the magnetic structure
on the $ac$ plane highlighting counter-rotation of moments between
the different sites (1-4,1$'$-4$'$) of the primitive cell.
Light/dark shading of the elliptical moment envelopes indicate an
alternating tilt of the plane of moments' rotation away from the
($ac$) plane. Right-hand labels (b)-(e) indicate where slices
through the magnetic structure are taken at different heights in
the unit cell and projected onto the ($ab$) plane to illustrate
the direction of the zig-zag chains. Note the alternating tilt of
the plane of rotation of the moments away from the $ac$ plane by
$\pm\phi$ between consecutive zig-zag chains stacked along $c$.
\label{fig:magstruct_full}}
\end{figure}

\begin{table}[tbh]
\caption{Fractional atomic coordinates of the iridium sites in the
primitive cell and corresponding magnetic basis vector components
in the determined magnetic structure.} \label{tab:magstruct}
\par
\begin{center}
\begin{tabular}
[c]{l|l|lll}\hline Site & Coordinates & $v_x$ & $v_y$ & $v_z$
\\\hline
 1 & $(0.25,0.25,z)$     & $+$ & $+$ & $+$\\
 2 & $(0.25,0.75,0.5-z)$ & $-$ & $+$ & $+$\\
 3 & $(0.25,0.25,1-z)$   & $-$ & $+$ & $+$\\
 4 & $(0.25,0.75,0.5+z)$ & $+$ & $+$ & $+$\\\hline
$1'$ & $(0.5,0.5,z')$    & $-$ & $-$ & $+$\\
$2'$ & $(0.5,0.5,0.5-z')$& $+$ & $-$ & $+$\\
$3'$ & $(0.5,0.5,1-z')$  & $+$ & $-$ & $+$\\
$4'$ & $(0.5,0.5,0.5+z')$& $-$ & $-$ & $+$\\\hline
\end{tabular}
\end{center}
\end{table}
The magnetic ions are located on the two iridium sublattices, Ir
at $8k$ $(0.25,0.25,z)$, $z=0.0836(2)\approx1/12$ and Ir$'$ at
$8i$ $(0.5,0.5,z')$, $z'=0.1670(3)\approx1/6$, each with four
sites in the primitive unit cell labelled $1-4$ and $1'-4'$ with
coordinates listed explicitly in Table~\ref{tab:magstruct} and
positions labelled in Fig.\ \ref{fig:magstruct_full}. For a
magnetic structure with propagation vector $\bm{q}=(q,0,0)$
symmetry analysis \cite{basireps} gives four types of basis
vectors $++++$ ($F$), $++--$ ($C$), $+--+$ ($A$) and $+-+-$ ($G$)
for each of the two iridium sublattices, which transform according
to the irreducible representations listed in
Table~\ref{tab:irreps}. The basis vectors encode symmetry-imposed
relations between the Fourier components of the magnetic moments
at the different sites, i.e. for basis vector $A$ on the Ir
sublattice the Fourier components on sites 1-4 are related by
$\bm{M}_{\pm\bm{q},1}$=$-\bm{M}_{\pm\bm{q},2}$=$-\bm{M}_{\pm\bm{q},3}$=$\bm{M}_{\pm\bm{q},4}$.
As described in the main text based on diffraction data the basis
vectors are found to occur in the combination
$i(A,-A)_x,i(-1)^m(F,-F)_y, (F,F)_z$ with moment amplitudes $M_x$,
$M_y$, $M_z$, and $m=1$ or 2. In both cases the magnetic structure
corresponds to a mixture of two irreducible representations
$\Gamma_3$ ($A_x$ and $F_z$) and $\Gamma_4(F_y)$ in
Table~\ref{tab:irreps}. The magnetic moment at position $\bm{r}$
belonging to a site index $n=1-4,1'-4'$ is
\begin{eqnarray}
\bm{M}_n(\bm{r}) & = & \left[ \bm{\hat{x}} M_x v_x(n) + (-1)^m
\bm{\hat{y}} M_y v_y(n) \right] \sin
\bm{q}\!\cdot\!\bm{r} \nonumber \\
&  + &\bm{\hat{z}} M_z v_z(n) \cos \bm{q}\!\cdot\!\bm{r}
\label{eq:Mr}
\end{eqnarray}
where $\bm{\hat{x}}$, $\bm{\hat{y}}$, $\bm{\hat{z}}$ are unit
vectors along the orthorhombic $\bm{a}$, $\bm{b}$, $\bm{c}$ axes,
respectively. $v_{x,y,z}$ are combined (8-site) basis vectors for
the two sublattices expressed in shorthand vector notation as
$v_x=(A,-A)$, $v_y=(F,-F)$, $v_z=(F,F)$ and with values listed
explicitly for all sites in the primitive cell in Table
\ref{tab:magstruct}. The Fourier components of the magnetic
moments are
\begin{eqnarray}
\bm{M}_{\bm{q},n} &= & i \left[\bm{\hat{x}} \frac{M_x}{2}
v_x(n)+(-1)^m \bm{\hat{y}} \frac{M_y}{2} v_y(n) \right] \nonumber \\
&+ &\bm{\hat{z}} \frac{M_z}{2} v_z(n)\label{eq:Mq}
\end{eqnarray}
with $\bm{M}_{-\bm{q},n} =\bm{M}^*_{\bm{q},n}$ as the magnetic
moment distribution is real. Eqs.\ (\ref{eq:Mr},\ref{eq:Mq})
describe all iridium sites, including those related by the
$C$-centering translation $(\frac{1}{2}\frac{1}{2}0)$, where
$\bm{r}$ is the actual position of the ion and $n$ is the site
index at the equivalent position
($1-4,1'-4'$) in the primitive unit cell.\\

\section{S3. Magnetic structure factors}

The structure factor for a magnetic Bragg reflection at wavevector
$\bm{Q}=(h,k,l) \pm \bm{q}$ is
\begin{equation}
\bm{\mathcal{F}}(\bm{Q})=
\bm{\mathcal{F}}((k,k,l){\pm}\bm{q})=\sum_{n}\bm{M}_{\pm
\bm{q},n}e^{i\bm{Q \cdot r}_n} \label{eq:structfact}
\end{equation}
where the sum extends over all 16 iridium sites in the
orthorhombic unit cell ($1$-$4$, 1$'$-4$'$ and their
$C$-translated positions) and $\bm{M}_{\pm \bm{q},n}$ are the
Fourier components of the magnetic moments at site $n$ with
position in the unit cell $\bm{r}_n$. Direct calculation of the
structure factors shows that the four types of basis vectors $F$,
$C$, $A$ and $G$ produce finite intensity magnetic peaks only for
satellites of certain integer $(h,k,l)$ reciprocal lattice
positions and not others. To make those selection rules
transparent we have calculated the structure factors analytically
for an ``ideal" iridium lattice, which is very close to the one
refined experimentally in Table\ \ref{tab:struc}, but with the Ir
coordinate at $z=1/12$ and Ir$'$ at $z'=1/6$. In this case the
structure factors are obtained as:
\begin{eqnarray}
\bm{\mathcal{F}}_F(\bm{Q})& = & 8 e^{i \zeta_{\pm}} \bm{M}_{\pm
\bm{q},1} \cos \frac{\pi l}{6} \left[
\cos \frac{\pi (k+l)}{2} \right.\nonumber \\
& & \quad \quad \left. + e^{\pm i \alpha} e^{i\frac{\pi(h+l)}{2}}
\cos\frac{\pi l}{2} \right] \delta_{h+k,2p} \label{eq:structF}
\end{eqnarray}

\begin{eqnarray}
\bm{\mathcal{F}}_C(\bm{Q})& = & 8e^{i \zeta_{ \pm}} \bm{M}_{\pm
\bm{q},1}\left[ \sin\frac{\pi l}{6} \sin \frac{\pi (k+l)}{2} \right.\nonumber \\
& & \left. - i e^{\pm i \alpha} e^{i\frac{\pi (h+l)}{2}} \cos
\frac{\pi l}{6} \sin \frac{\pi l}{2} \right] \delta_{h+k,2p}
\label{eq:structC}
\end{eqnarray}

\begin{eqnarray}
\bm{\mathcal{F}}_A(\bm{Q})& = & 8ie^{i \zeta_{ \pm}}\bm{M}_{\pm
\bm{q},1}\sin \frac{\pi l}{6} \left[ \cos \frac{\pi (k+l)}{2} \right.\nonumber \\
& & \quad \quad \left. - e^{\pm i \alpha} e^{i\frac{\pi (h+l)}{2}
} \cos \frac{\pi l}{2} \right] \delta_{h+k,2p} \label{eq:structA}
\end{eqnarray}

\begin{eqnarray}
\bm{\mathcal{F}}_G(\bm{Q})&= &-8e^{i \zeta_{ \pm}}\bm{M}_{\pm
\bm{q},1}\left[ i\cos\frac{\pi l}{6} \sin \frac{\pi (k+l)}{2}  \right.\nonumber \\
& & \left. + e^{\pm i \alpha} e^{i\frac{\pi (h+l)}{2}} \sin
\frac{\pi l}{6} \sin \frac{\pi l}{2} \right] \delta_{h+k,2p}
\label{eq:structG}
\end{eqnarray}
where $\delta$ is the Kronecker symbol and $p$ is an integer, i.e.
\begin{equation} \delta_{h+k,2p} = \left\{
\begin{array}{l l}
1 , & \quad h+k ~ \text{even}\\
0 , & \quad h+k ~ \text{odd} \nonumber
\end{array} \right.
\label{eq:delta}
\end{equation}
and this term arises in the structure factor due to the
$C$-centering. Here $\zeta_{ \pm}=\pi(h/2 \pm q/2 +k+l/3)$ and the
1st and 2nd terms in the square brackets come from the Ir and
Ir$'$ sublattices, respectively, where we have assumed that their
magnetic Fourier components are the same up to a complex phase
offset $\alpha$, i.e.
$\bm{M}_{\bm{q},1'}=e^{i\alpha}\bm{M}_{\bm{q},1}$. In the
determined magnetic structure the phase offset for the $A_x$ and
$F_y$ basis vectors is $\alpha=\pi$, whereas for the $F_z$ basis
vector it is $\alpha=0$.

From the above equations it is clear that in the ($h0l$) plane,
for $l=6n$ ($n$ integer) only $F$-basis vectors contribute as the
structure factor for all the other basis vectors cancels
($\mathcal{F}_C=\mathcal{F}_A=\mathcal{F}_G=0$). In this case
further inspection of the structure factors shows that satellites
can be separated into those corresponding to $\alpha=0$ and those
with $\alpha=\pi$, for example the magnetic satellites at
$(0,0,24)\pm\bm{q}$ come from an $(F,F)$ basis vector ($\alpha=0$)
and peaks at $(2,0,24)\pm\bm{q}$ come from an $(F,-F)$ basis
vector ($\alpha=\pi$). For the $(h,0,l)$ plane depicted in Fig.\
\ref{fig:hscan}b) magnetic satellite peaks occur only for $h$ even
with the further selection rule $l$ odd for both $C$ and $G$, and
$l$ even for both $F$, $A$ with pure $F$ (no $A$ contribution) for
$l=6n$, $n$ integer. Satellites of (odd, odd, $l=3+6n$), $n$
integer such as (1,1,21) are of $A$, $G$ type (no $F$ or $C$
contribution) and this is used to prove the existence
of an $A$-basis vector in the magnetic ground state.\\

\section{S4. Resonant magnetic x-ray diffraction experiments}

The sample used in the x-ray experiments was a single crystal of
$\gamma$-Li$_2$IrO$_3$ with a typical rhombic morphology
\cite{modic} of volume $\sim\,$35$\times10^3\mu$m$^3$. The sample
quality was checked using a Mo-source SuperNova x-ray
diffractometer confirming the previously deduced crystal structure
(orthorhombic space group $Cccm$ with lattice parameters
$a=5.9119$~\AA, $b=8.4461$~\AA, $c=17.8363$~\AA~ at room
temperature). Specific heat measurements on this crystal using an
in-house ac micro-calorimeter observed a clear anomaly near
$T_{\rm N}=39.5$\ K [see Fig.~\ref{fig:braggpeak}b) inset], in
good agreement with the transition temperature to magnetic order
inferred from earlier magnetic susceptibility and torque
measurements \cite{modic}.

Resonant x-ray diffraction at the Ir L$3$ edge (11.215\ keV) was
performed using the I16 beamline at Diamond in reflection geometry
with the crystal mounted with the (001) axis surface normal. The
$\sigma$-polarized incident beam was de-focused to an area
$200\times200\,\mu$m$^2$, to ensure illumination of the entire
sample. For the polarization analysis measurements a Au ($3,3,3$)
crystal was placed in the scattered beam and intensities were
counted in an APD detector, for the rest of the measurements an
area detector (Pilatus) was used. The sample was cooled using a
closed-cycle refrigerator (CCR) with Be windows with a base
temperature of 9\ K.

\section{S5. Intensity in resonant magnetic x-ray diffraction}

In the dipolar approximation the magnetic resonant x-ray
scattering intensity is proportional to
\begin{equation}
L(\theta) \mathcal{A} \left|(\bm{\hat{\epsilon'}} \times
\bm{\hat{\epsilon}}) \cdot\bm{\mathcal{F}}(\bm{Q})\right|^2
\nonumber 
\end{equation}
where $L(\theta$) is the Lorentz factor at the scattering angle
$2\theta$, $\mathcal{A}$ is an absorption correction dependent
upon the experimental geometry, $\bm{\mathcal{F}}(\bm{Q})$ is the
magnetic structure factor given in eq.\ (\ref{eq:structfact}), and
$\bm{\hat{\epsilon'}}$ and $\bm{\hat{\epsilon}}$ are unit vectors
along the polarization of the electric field component of the
scattered and incident x-ray beams, respectively \cite{hill}. For
a $\sigma$-polarized incident beam magnetic resonant scattering
occurs only in the $\sigma$-$\pi'$ channel [see diagram in Fig.\
\ref{fig:azimuths}a) inset], meaning that the product of the
electric field polarization vectors is along the scattered beam
direction, i.e. $\bm{\hat{\epsilon'}} \times
\bm{\hat{\epsilon}}=\bm{\hat{k'}}$.

\begin{figure}[htbp]
\includegraphics[width=0.48\textwidth]{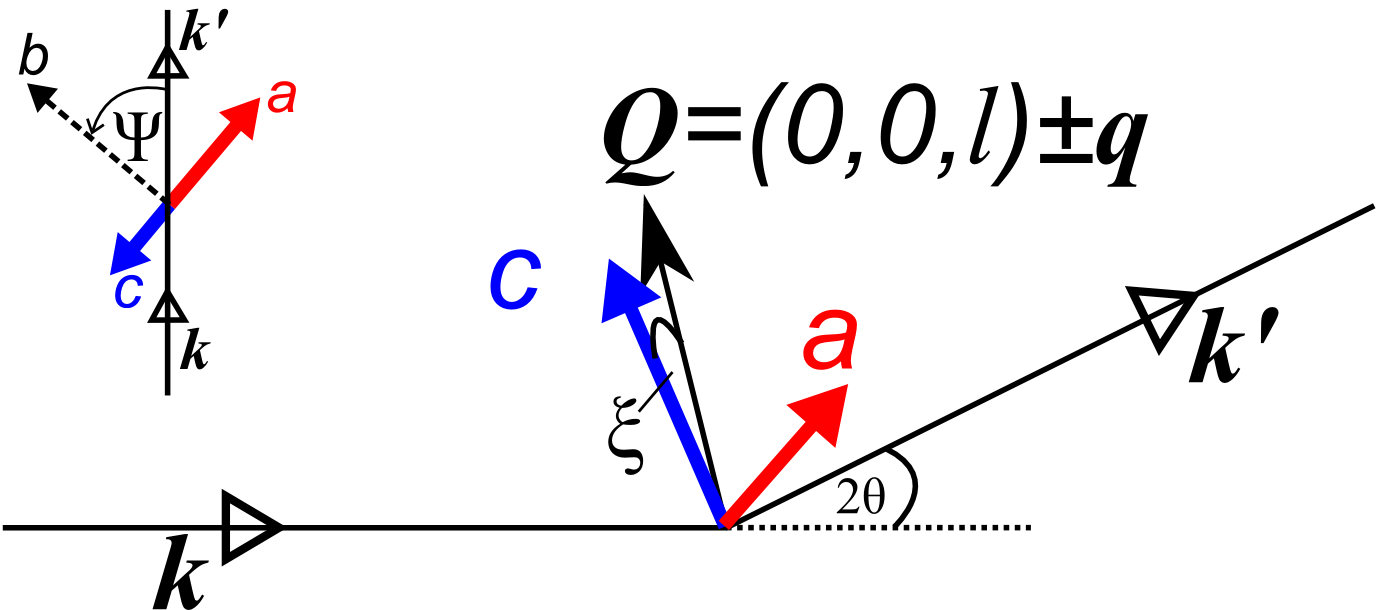}
\caption{(color online) Schematic of the x-ray scattering
experiment probing the magnetic scattering at $\bm{Q}=(0,0,l) \pm
\bm{q}$ showing the orientation of the orthorhombic crystal axes
at a general azimuth angle $\Psi$ (inset shows projection in the
plane normal to $\bm{Q}$ indicating the azimuth angle origin). In
the main diagram $\xi$ is the angle between $\bm{Q}$ and the
$c$-axis, i.e. $\xi=\cos^{-1}( \bm{\hat{Q}} \cdot \bm{\hat{c}}
)$.} \label{fig:expsetup}
\end{figure}

The orientation of the magnetic moments can be experimentally
determined by exploiting the polarization dependence of the
scattering intensity as only the component of the structure factor
vector along the scattered beam direction,
$\mathcal{F}_{\parallel}=\bm{\mathcal{F}}\cdot\bm{\hat{k'}}$,
contributes. By keeping the instrument in the scattering condition
and rotating the sample around the scattering wavevector
$\bm{Q}=\bm{k'}-\bm{k}$ the projection $\mathcal{F}_{\parallel}$
of the structure factor vector varies depending on the azimuth
angle $\Psi$ with maximum intensity when the moments that give
rise to the scattering make the smallest angle with
$\bm{\hat{k'}}$ and zero intensity when they are perpendicular.
This is illustrated by the calculated azimuth dependence of the
intensity for moments along $x$, $y$ and $z$ for an $A$-type basis
vector in Fig.\ \ref{fig:azimuths}a) (red, blue, green curves),
and the data uniquely identifies that scattering comes from a
basis vector with collinear magnetic moments along $x$.

For magnetic Bragg reflections where moments along two orthogonal
directions contribute, such as $(0,0,16)+\bm{q}$ in Fig.\
\ref{fig:azimuths}c) of mixed $A_x$, $F_z$ character, the
scattering intensity has a cross term that is sensitive to the
complex phase between the structure factor vectors along the two
directions, so it can distinguish between moments varying
sinusoidally along a line in the $xz$ plane (phase offset 0 or
$\pi$) or rotating in the $xz$ plane (phase offset $\pm pi/2$).
Explicitly, the parallel component of the structure factor when
both $x$ and $z$ moments contribute is
\begin{eqnarray}
\mathcal{F}_{\parallel}(\Psi)&=&\mathcal{F}_z(\cos\xi\sin\theta-\sin\xi\cos\theta\sin\Psi)\nonumber \\
& & +\mathcal{F}_x(\sin\xi\sin\theta+\cos\xi\cos\theta\sin\Psi)
\nonumber 
\end{eqnarray}
where $\mathcal{F}_{x,z}$ are the separate structure factors for
the magnetic moment components along the $x,z$-axes and the angles
$2\theta$ and $\xi$ are defined in the scattering diagram in Fig.\
\ref{fig:expsetup}. From this the azimuth dependence of the
magnetic scattering intensity is obtained as
\begin{eqnarray}
|\mathcal{F}_\parallel(\Psi)|^2&=&|\mathcal{F}_z|^2(\cos\xi\sin\theta-\sin\xi\cos\theta\sin\Psi)^2 \nonumber \\
& & +|\mathcal{F}_x|^2(\sin\xi\sin\theta+\cos\xi\cos\theta\sin\Psi)^2 \nonumber \\
& &  + {\rm{Re}}(\mathcal{F}_z {\mathcal{F}_x^*})\left[\sin 2\xi \left( \sin^2 \theta-\cos^2 \theta \sin^2 \Psi\right) \right. \nonumber \\
& & \quad \quad \quad \quad \quad \quad \quad \quad + \left. \sin
2 \theta\cos 2\xi\sin\Psi\right] \label{eq:structfull}
\end{eqnarray}
where $\rm{Re}()$ means the real part. The first two terms give
the sum of the scattering intensities separately from moments
along the two directions, whereas the last term is the cross term.
For the specific case of $\bm{Q}=(0,0,l$=$16)+\bm{q}$, the cross
term pre-factor is
\begin{equation}
{\rm{Re}}(\mathcal{F}_z\mathcal{F}_x^*) \propto \left\{
\begin{array}{r l}
\mp\cos\frac{\pi l}{6} \sin\frac{\pi l}{6}, & \pm i(A,-A)_x, (F,F)_z\\
0, &  \pm(A,-A)_x, (F,F)_z
\end{array} \right.\nonumber
\label{eq:structxz}
\end{equation}
where $\mathcal{F}_x$ and $\mathcal{F}_z$ are of the form
$\mathcal{F}_{(A,-A)}$ and $\mathcal{F}_{(F,F)}$, respectively,
from eqs.\ (\ref{eq:structF},\ref{eq:structA}). There is no
cross-term if moments are in-phase or $\pi$ out-of-phase and a
finite cross-term if moments are $\pm \pi/2$ out-of-phase. The
azimuth data in Fig.\ \ref{fig:azimuths}c) clearly shows a large
asymmetry around $\Psi=0$, which is quantitatively explained (red
line) by the basis vector combination $i(A,-A)_x, (F,F)_z$ with
moments rotating in the $xz$ plane and rules out in-phase or $\pi$
out-of-phase moments along the $x$ and $z$ axes (blue line).

\section{S6. Magnetic structure with counter-rotating moments stabilized by Kitaev interactions}

In this section we give details of the derivation of the net spin
correlation between nearest-neighbor sites along the $d$-bonds
$\left\langle S_n^\mathsf{\eta} S_{n'}^\mathsf{\eta}
\right\rangle_{\mathsf{\eta}} = \langle S^a \rangle \langle S^c
\rangle \tfrac{1}{2} \sin \tfrac{\pi q}{2}$ with $\eta=\mathsf{x}$
or $\mathsf{y}$. First we recall the transformation from the
crystallographic axes $\bm{\hat{a}},\bm{\hat{b}},\bm{\hat{c}}$ to
the Kitaev axes defined as $\mathsfbf{\hat{x}} =
(\bm{\hat{a}}+\bm{\hat{c}})/\sqrt{2}$,
$\mathsfbf{\hat{y}}=(\bm{\hat{a}}-\bm{\hat{c}})/\sqrt{2}$ and
$\mathsfbf{\hat{z}}= \bm{\hat{b}}$, see Fig.\ \ref{fig:magstruct}.

The $a$-component of the displacement between adjacent sites of
type 1 and 1$'$ in a zig-zag chain is
$(\bm{r}_1-\bm{r}_{1'})\!\cdot\!\bm{\hat{a}} = \pm a/4$, where the
upper (lower) sign is to be taken if the two sites are connected
by a Kitaev bond of type $\mathsf{x}$ ($\mathsf{y}$). More
generally, for neighboring sites of type $n$ and $n'$ the
displacement projection is $(\bm{r}_n-\bm{r}_{n'} )\cdot
\bm{\hat{a}} = \pm \nu_n a/4$, where $\nu_n = +1$ for $n=1,4$ and
$\nu_n = -1$ for $n=2,3$.

In this notation, we obtain from (\ref{eq:Mr}) that sites in the
rotating magnetic structure carry the spin moment
\begin{eqnarray} \bm{S}_{n,n'}(\bm{r}) & = & \pm
 \left[ \nu_n \bm{\hat{a}} \langle S^a \rangle   +(-1)^m
 \bm{\hat{b}}\langle S^b\rangle   \right]
\sin \bm{q}\cdot \bm{r} +\nonumber \\
& & \bm{\hat{c}}\langle S^c \rangle
 \cos \bm{q} \cdot\bm{r}
\label{eq:Sr} \end{eqnarray} where the $\pm$ sign in front of the
square bracket corresponds to unprimed/primed sites and the case
$m=1$ is depicted in Fig.\ \ref{fig:magstruct_full}. The product
of this $\pm$ sign in front of the square brackets and the $\nu_n$
sign factor gives a sign which alternates between $+$ and $-$ when
sites are listed by their $c$-coordinate, i.e. the vertical axis
in Fig.\ \ref{fig:magstruct}, producing the counter-rotation of
the spin moments in the $ac$ plane.

It is immediately evident that along Kitaev $\mathsf{z}$-type
bonds (linking sites of type $1'2'$, $3'4'$, $13$ and $24$, see
Fig.\ \ref{fig:magstruct_full}a)) the $S^{\mathsf{z}}$$=$$S^{b}$
spin components are always ferromagnetically-correlated, enabling
energetic stabilization through the strong FM Kitaev interaction
on these bonds, $K_c<0$. The more subtle correlations, as
discussed in the main text, are those of the
$S^{\mathsf{x}}$($S^{\mathsf{y}}$) spin components across
$\mathsf{x}$-type ($\mathsf{y}$-type) Kitaev bonds. The
counter-rotation of neighboring moments within the unit cell
enables these subtle Kitaev correlations, as follows
\begin{align}
& \left\langle S_{n}^{\eta} S_{n'}^{\eta} \right\rangle_{\bm{r}: \
\eta \rm{-bond}} /\langle S^a \rangle /\langle S^c
\rangle  = \nonumber\\
& \left\langle
 \cos\left(\bm{q}\!\cdot\! \bm{r} \pm \nu_n \frac{\bm{q}\!\cdot\!\bm{a} - \pi}{4}\right)
  \cos\left(\bm{q}\!\cdot\!\bm{r} \pm \nu_n \frac{\pi}{4}\right)
 \right\rangle_{\bm{r}}
   = \nonumber\\
& \frac{1}{2}  \cos\left(\frac{\bm{q}\!\cdot\!\bm{a} -
2\pi}{4}\right) = \frac{1}{2}  \sin\left(\frac{\bm{q}\!\cdot\!
\bm{a}}{4}\right)
\end{align}
with $\eta=\mathsf{x}$ or $\mathsf{y}$ and $\langle \ldots
\rangle_{\bm{r}}$ indicates the average over all positions
$\bm{r}$ of sites of type $n$ in the crystal. Note that defining
the rotating magnetic structure within the primitive unit cell
(containing 8 sites) is sufficient to uniquely specify the spin
moments on all sites in the crystallographic $a,b,c$ unit cell,
which contains 16 iridium sites; here $\bm{q}\cdot \bm{a}$ ranges
from $-2\pi$ to $2\pi$. Within our convention of the spin
components within the unit cell, positive values of $q$ (i.e. $0<
\bm{q}\cdot \bm{a} < 2\pi$) correspond to positive Kitaev
correlations, which may be stabilized by FM Kitaev interactions
($K_d<0$).

\section{S7. Luttinger-Tisza analysis of the minimal model Hamiltonian}

We diagonalize the spin Hamiltonian in momentum space without the
unit length constraint. The energies and modes are found as the
eigenstates of the $24 \times 24$ matrix, corresponding to three
spin components for each site in the primitive unit cell. Then
solutions obeying the unit length constraint are constructed from
the lowest eigenmode, possibly with higher energy modes mixed in.

The lowest eigenvalue of the Hamiltonian in eqn.\ (\ref{eq:ham})
with parameters as given in the main text occurs at a wavevector
numerically indistinguishable from $(4/7,0,0)$ (in r.l.u's of the
orthorhombic unit cell $a \times b \times c$). This minimal energy
eigenmode, with energy $-13.6$ meV, has the ordered spin moment
$\bm{S} \propto \bm{\hat{c}} \pm i 0.85 \nu_n\bm{\hat{a}}$, with
the upper (lower) sign for the unprimed (primed) sites, and hence
does not quite obey the constraint of normalized spins. However it
does exactly describe the coplanar projection of the experimental
magnetic structure onto the $ac$ plane. The next three eigenmodes
again involve only $S^a,S^c$ spin components, and cannot mix with
the lowest mode. The fifth eigenmode at this wavevector, with
energy $-10.5$ meV, has spins purely along $\bm{\hat{b}}$, with an
order pattern of $\pm$ signs for unprimed/primed sites, exactly
capturing the pattern of the non-coplanar tilts in the
experimentally-determined structure. So mixing between this
eigenmode and the lowest energy eigenmode to ensure the constraint
of fixed-length spins can match all features of the
experimentally-determined magnetic structure. For completeness we
note that changing the sign of the mixing coefficient corresponds
to changing between the cases $m=1$ and $2$ in eq.\ (\ref{eq:Sr}),
with the two structures being degenerate in energy.

In summary, through extensive searches in parameter space for
candidate spin Hamiltonians we have found that all couplings in
eqn.\ (\ref{eq:ham}) are required to stabilize the observed
magnetic structure as the lowest-energy structure with
fixed-length spin moments. The phase obtained is stable within a
range of values for the Hamiltonian parameters and the quoted
values in the text are a representative solution, where the
overall scale is set by the constraint that the calculated
transition temperature to magnetic order matches the experimental
value.

\begin{thebibliography}{99}
\bibitem{error} The uncertainty in the value of $q$ is estimated
from combining uncertainties in the crystal orientation
(UB-matrix) and the I16 instrument intrinsic sphere of
uncertainty.
\end{thebibliography}
\end{document}